# Emerging organic contaminants in wastewater: Understanding electrochemical reactors for triclosan and its by-products degradation

C. Magro[1*], E.P. Mateus[1], J.M. Paz-Garcia[2], & A.B. Ribeiro[1*]

[1]CENSE, Department of Sciences and Environmental Engineering, NOVA School of Science and Technology, NOVA University Lisbon, Caparica Campus, 2829-516 Caparica, Portugal

[2]Department of Chemical Engineering, Faculty of Sciences, University of Malaga, Teatinos Campus, 29010 Málaga, Spain

* Correspondence: c.magro@campus.fct.unl.pt; abr@fct.unl.pt

## Abstract

Degradation technologies applied to emerging organic contaminants from human activities are one of the major water challenges in the contamination legacy. Triclosan is an emerging contaminant, commonly used as antibacterial agent in personal care products. Triclosan is stable, lipophilic and it is proved to have ecotoxicology effects in organics. This induces great concern since its elimination in wastewater treatment plants is not efficient and its by-products (e.g. methyl-triclosan, 2,4-dichlorophenol or 2,4,6-trichlorophenol) are even more hazardous to several environmental compartments. This work provides understanding of two different electrochemical reactors for the degradation of triclosan and its derivative by-products in effluent. A batch reactor and a flow reactor (mimicking a secondary settling tank in a wastewater treatment plant) were tested with two different working anodes: Ti/MMO and Nb/BDD. The degradation efficiency and kinetics were evaluated to find the best combination of current density, electrodes and set-up design. For both reactors the best electrode combination was achieved with Ti/MMO as anode. The batch reactor at 7 mA/cm$_2$ during 4 h attained degradation rates below the detection limit for triclosan and 2,4,6-trichlorophenol and, 94% and 43% for 2,4-dichlorophenol and methyl triclosan, respectively. The flow reactor obtained, in approximately 1 h, degradation efficiencies between 41% and 87% for the four contaminants under study. This study suggests an alternative technology for emergent organic contaminants degradation, since the combination of a low current density with the flow and matrix induced disturbance increases and speeds up the compounds' elimination in a real environmental matrix.

Abbreviations: BDD - Boron-doped diamond; DCP - 2,4-dichlorophenol; EBR - Electrochemical Batch Reactor; EFR - Electrochemical Flow Reactor; EOCs - emerging organic contaminants; MTCS - methyl-triclosan; MMO - mixed metal oxides; TCP - 2,4,6-trichlorophenol; TCS - triclosan.

## 1. Introduction

The world's rapid population growth over the last century has been a major factor into the demand for water resources usage and reuse. To overcome these water challenges, water contamination must be taken into account. Recently, the environmental quality criteria of water resources have been linked to a new class of environmental pollutants, the emerging organic contaminants (EOCs), a reality that has increased the need for sustainable tools that guarantee their quality and safety standards, enable their monitoring and promote the prosperity of a healthy population and environment (Corcoran et al., 2010). EOCs are defined as "chemical substances that have no regulation and are suspected to negatively affect the environment or whose effects are unknown" (Daughton, 2004; Geissen et al., 2015). Among EOCs, triclosan (TCS, 2,4,4'-Trichloro-2'-hydroxydiphenyl ether) is an antimicrobial agent that has been used for more than 50 years as an antiseptic, disinfectant or preservative in clinical settings and several consumer products. TCS has been detected in wastewater treatments plants (Brose et al., 2019; Chen et al., 2019; Halden, 2019) and in surface water (Hua et al., 2005; McAvoy et al., 2002). Recent reviews on TCS recount numerous health effects ranging from endocrine-disruption to uncoupling mitochondria (Olaniyan et al., 2016; Weatherly and Gosse, 2017). Although in 2016 the US Food and Drug Administration banned TCS from certain wash products, namely hand soap and body wash (Food and Drug Administration, 2016) and hospital products by the end of 2018 (Food and Drug Administration, 2017), it is permissible to have TCS in e.g. toothpastes, cosmetics, clothes or toys (Bever et al., 2018).

Furthermore, TCS derivates, such as the metabolite methyl-triclosan (MTCS) (Guo et al., 2009), which is even more persistent (Balmer et al., 2004) and TCS by-products, formed by repeated exposure to chlorine in water such as 2,4-dichlorophenol (DCP) and 2,4,6-trichlorophenol (TCP) are also under concern, since they present health risk to humans and are recognized as persistent priority pollutants in the United States, Europe and China (Xing et al., 2012).

Several clean-up technologies have been developed and used to improve the quality and safety of water reuses. Among these technologies, the electrochemically-induced advanced oxidation processes (e.g. Fenton's reaction and anodic oxidation) have been receiving special attention (Glaze et al., 1987; Oturan and Aaron, 2014; Panizza and Cerisola, 2009). Another alternative is the electrokinetic process which is based on the application of a low-level direct current between a pair of electrodes, that in addition remove contaminants from the contaminated matrix by electric potential also promotes the generation of •OH and therefore enhances the oxidation of EOCs. This technology proved to be efficient in the degradation of EOCs in soil (Guedes et al., 2014), sludge (Guedes et al., 2015) and effluent (Ferreira et al., 2018). The optimal combination between the electrode materials (e.g. titanium/mixed metal oxides (Ti/MMO), boron-doped diamond (BDD), platinum) and the reactor design are key factors, since the oxidation process is dependent on the materials nature and the reactor workability (Schranck and Doudrick, 2020; Walsh and Ponce de León, 2018). Regarding the reported works on different electrode materials and batch/flow reactors, Ren et al. (2016) presented a vertical-flow electro-Fenton reactor, composed of 10 cell compartments using $PbO_2$ anode and modified graphite felt mesh cathode for the degradation of tartrazine, reaching with the optimal conditions, TOC removal efficiency of 100%. Pérez et al. (2017) studied a microfluidic flow-through electrochemical reactor for wastewater treatment, that achieved, with diamond anodes, complete mineralization of clopyralid spiked in a low-conductive matrix. Wang et al. (2019) reported a continuous-flow reactor for electrochemical oxidation of various alcohols using carbon anode, where with 800 mA enabled effective oxidation up to 99% yield in 10 min. Moreover, comparative studies in electrodes combinations were described: Yoon et al. (2012) reported a flow reactor for the electrochemical degradation of phenol and 2-chlorophenol using Pt/Ti and BDD electrodes, as well as Ambauen et al. (2019) comprised a electrochemical oxidation batch reactor for salicylic acid degradation with BDD and Pt electrodes. In both studies similar removal rates in the different electrodes combinations were attained, showing that not only the electrodes type highly influence the compounds degradation efficiencies, but also the physicochemical characteristics of the contaminants to be degraded. BDD and MMO have been mainly and equally used as anodes (Moreira et al., 2017), both showing similar performances in the degradation efficiency (Brillas and Martínez-Huitle, 2015; Skoumal et al., 2008; Yoon et al., 2012). BDD was reported as electrochemical inactivator of phenolic compounds (Sirés et al., 2007; Wang and Farrell, 2004), and Ti/MMO was used to degrade organic contaminants in wastewater (Yuan et al., 2013).

The aim of this study was to carry out an experimental electrochemical treatment for the degradation of TCS and its byproducts MTCS, DCP and TCP, in a real wastewater matrix: a secondary effluent. A batch reactor was the starting core, with experiments on (1) electrodes combination (Ti/MMO as anode and cathode; BDD/Nb as anode and Ti/MMO as cathode); (2) current densities; (3) degradation kinetics. Furthermore, a flow reactor, designed to mimic a secondary settling tank in a wastewater treatment plant, was introduced to find a system that has the potential for operational implementation.

## 2. Materials and Methods

### 2.1 Chemical, standards and effluent characteristics

TCS (99%), MTCS (99%), DCP (98%) and TCP (98%) were purchased from Sigma–Aldrich (Steinheim, Germany), see Table S1. Individual stock solutions for calibration purposes were prepared with 1000 mg/L in methanol and stored at −18 °C. The methanol, acetonitrile, acetone and formic acid used were from Sigma–Aldrich (Steinheim, Germany) in gradient grade type. Water (Type I) was from a Millipore system (Bedford, MA, USA). The effluent used was the liquid fraction collected in the secondary settling tank at a wastewater treatment plant (Lisbon, Portugal), which initial characterization is presented in Table 1.

## 2.2 Methods

### 2.2.1 Electrochemical degradation reactors

Two different set-ups are represented in Figure 1. The experiments were carried out in an electrochemical batch reactor (EBR) and in an electrochemical flow reactor (EFR). The electrodes used were made of Ti/MMO Permaskand wire: $Ø = 3$ mm, $L = 80$ mm (Grønvold & Karnov A/S, Denmark) and Nb/BDD plate: $H = 50$ mm, $L = 80$ mm, $T = 1$ mm (Neocoat, Switzerland), assembled in different configurations (as anode or as cathode). A power supply E3612A (Hewlett Packard, Palo Alto, USA) was used to maintain a constant current in the electrochemical reactors and the voltage was monitored. The volume of effluent treated was approximately 500 mL for each reactor. The effluent flow (9 mL/min) in the EFR was maintained by a peristaltic pump (Watson-Marlow 503 U/R, Watson-Marlow Pumps Group, Falmouth, Cornwall, UK).

All electrochemical experiments were carried out in duplicate, in dark conditions, and at a controlled room temperature of 22 °C, according to the conditions presented in Table 2. Control experiments, without applying current, were also carried out. During the experiments, the pH was measured with a Radiometer pH-electrode EDGE (HANNA, USA) and the conductivity with a Radiometer Analytic LAQUA twin (HORIBA Ltd., Japan). In both reactors (EBR and EFR) the matrix was spiked with the four compounds under study (TCS, MTCS, DCP and TCP), in order to monitor their degradation process (0.8 mg/L for each). To assess EOCs removal kinetics in the EBR, samples were collected every 15 min, during 4 h. For EOC determination and quantification, initial and final samples were extracted following the procedures described in section 2.2.2.

### 2.2.2 Instruments and analytical procedures

The effluent initial characterization was performed for the following chemical parameters: Chemical Oxygen Demand, COD ( determined by volumetric method after a previous oxidation with potassium dichromate in an open-reflux, at 160ºC, in an acidic environment, for 110 min); biological oxygen demand, $BOD_5$ (determined by dilution method and determination of the dissolved oxygen by using a specific probe: OxiTop IS6, GlobalW Gold River, CA, USA); Total P concentration were determined by Inductively Coupled Plasma with Optical Emission Spectrometry, ICP-OES (HORIBA Jobin-Yvon Ultima, Japan), equipped with generator RF (40.68 MHz), monochromator Czerny-Turner with 1.00 m (sequential), automatic sampler AS500 and dispositive CMA-Concomitant Metals Analyzer. $Cl^-$, $NO_3^-$ and $SO_4^{2-}$ were analyzed by Ion Chromatography, IC (DIONEX ICS-3000, USA), equipped with conductivity detector.

The extraction of the analytes in the effluent was performed by solid-phase extraction using Oasis HLB (200 mg, 6 mL; Waters Corporation, Saint-Quentin En Yvelines Cedex, France). The solid-phase extraction cartridges were conditioned by washing with 3 × (6 mL) of methanol, followed by re-equilibrium with 3 × (6 mL) of Milli-Q water. For organic compounds enrichment, the samples were acidified to pH = 2 before extraction using nitric acid. The 200 mL aqueous samples were passed through the cartridge at a flowrate of approximately 10 mL/min by applying a moderate vacuum, followed by a dried period of approximately 3 min by vacuum. The retained analytes were eluted sequentially with 2 × (4 mL) of methanol and 1 × (4 mL) of acetone, conferring a concentration factor of 16.7x to the analysis.

Determination of the target compounds was performed in an Agilent 1260 Infinity II high-performance liquid chromatography (HPLC) equipped with a quaternary pump and auto-sampler, and a diode array detector (DAD)/fluorescence detector 1100 Series (Agilent Technologies Inc., USA). The EC-C18 column (InfinityLab Poroshell 120 High Efficiency, 100 mm × 4.6 mm; 2.7 micron with Column ID, Agilent Technologies Inc., USA) was used. All HPLC runs were performed at a constant flow (1.5 mL/min), in gradient mode, with the oven set to 36 °C. A mixture of acetonitrile/Milli-Q water/formic acid was used as eluents (A: 5/94.5/0.5% and B: 94.5/5/0.5%) with a gradient of 60% of B (0–2 min) followed by 97% of B (2-3.5 min) and 98% of B until 5 min. Calibration curve was performed in the range between 0.5 and 20.0 mg/L. The limits of detection and quantification in this work were, respectively, 0.7 and 2.0 mg/L for TCS, 1.3 and 3.9 mg/L for MTCS, 0.7 and 2.0 mg/L for DCP, 1.0 and 3.0 mg/L for TCP. Recovery tests were made with fortified effluent for 1 h of contact time (30 min of slow agitation). The recovery percentages were between 62 and 120% in all cases. The same HPLC system was used to monitor the EOCs' degradation rates and kinetics behavior. The percentage of degradation was calculated according to Equation 1:

$$(1 - \frac{Final\ EOC\ mg/L}{Initial\ EOC\ mg/L} \times 100). \tag{1}$$

The mineralization of EOCs was analyzed from the decay of their total organic carbon, TOC, which was determined on a Vario TOC select analyzer (Elementar, Langenselbold, Germany), after filtration over 0.70 μm membrane filters. The screen and identification of new peaks related to new or related compounds was performed by a LC/MS, LC Agilent 1200 Series with Binary pump - MS Agilent 6130B Single Quadrupole (Agilent Technologies Inc., USA).

All sample analysis was carried out in duplicate. The statistical data obtained was analyzed by the GraphPad Prism version 8.0e. Statistically significant differences among samples for 95% level of significance were calculated through t-test on way and ANOVA.

## 3. Results and Discussion

In the following section, electrochemical reactors for the degradation of EOCs in effluent were tested (cases according to Table 2 conditions), correspondingly:

  i. EBR was tested to find the best current for both electrodes' combination: Cases 1 and 2;

  ii. EBR degradation kinetics was studied for the most appropriated current, for both electrodes' combination: Cases 1 and 2; electro-by products were investigated as well.

  iii. EFR degradation was studied for both electrodes combination: Cases 1 and 2, in order to have a final decision on which one is the best option for a reactor operational implementation.

## 3.1 Electrochemical treatment evaluation

### 3.1.1 Electrochemical Batch Reactor (EBR): pH, conductivity and voltage

Table 3 presents the pH, conductivity and voltage measured at the beginning and after 4 h of the EBR treatment.

In the control experiments (data not shown) there was no significant pH and conductivity changes in the effluent. pH with the Ti/MMO as anode remained approximately constant, whereas for Nb/BDD as anode, acidification occurred, as the applied current increased (initial vs final at different pH: $p$ value = 0.001-0.01 at 95% of confidence level). This indicates that Nb/BDD anode has a better tendency for water oxidation promoting, the acidification of the media (pH decreases from around 8 to 3.8).

The conductivity of the matrix increased in the experiment with the highest current density and only when using Nb/BDD as anode. The degradation process may cause the presence of more ions in solution, and consequently higher conductivity at the end of the experiments. The voltage drop behavior is in accordance with the changes observed in the conductivity. A slight variation of voltage between the working electrodes was observed in all cases, except in the control assay, related to the changes in conductivity due to electrolysis reactions. Concentration profiles within the reactor were observed, indicating that mixing could play an effect on energy requirements.

### 3.1.2 Electrochemical Batch Reactor (EBR): current density

Figure 2 shows the removal of the target EOCs, measured as the percentage of degradation. Control experiments showed no degradation. As the experiments were carried out in dark conditions, photodegradation was not taking place. Compound volatilization from effluent is not expected to be an important degradation process based on the estimated Henry's law constant of the studied compounds ($k_H > 10^{-6}$) (Table S1 at supplementary data). Thus, the degradation rates are attributed namely to (1) biotic factors, (2) direct anodic oxidation, and (3) indirect oxidation in the liquid bulk. Direct anodic oxidation occurs as the contaminants get in contact with anode surface and destroyed by the electron transfer reaction. Indirect oxidation in the liquid bulk is mediated by the oxidants that are formed electrochemically, such as chlorine, hypochlorite, hydroxyl radicals, ozone and hydrogen peroxide (Klavarioti et al., 2009). Also, if the pH during the experiments is greater than EOC's *pKa* (Ti/MMO 7 and 14 mA/cm$_2$), the compounds were in their ionized form, more susceptible to degradation (Chianese et al., 2016).

In a simpler design as EBR (see Figure 1), the main removal mechanism is the electro-degradation that corresponds to compounds' oxidation and reduction. At the anode, the EOCs suffers both direct and indirect degradation whereas, in the cathode, only indirect degradation will occur. As mentioned before, in the case 2 experiments (4, 8.5, 10 mA/cm$_2$, Nb/BDD as anode) the pH turned acidic, thus, the hydroxyl radical had a standard reduction potential of 2.8 V ( $^{\bullet}OH + e^- + H^+ = H_2O$ ). In the case 1 experiments the alkaline pH decreases the standard reduction potential to 1.8 V ( $^{\bullet}OH + e^- = OH^-$). Moreover, chemical structures (Table S1 at supplementary data) and reactivity will play a role in the degradation process. For instance, the phenolic ring of TCS can be activated by the two O-containing groups and may be attacked by •OH radicals, with production of hydroxylated TCS (Yuan et al., 2013). In both cases 1 and 2, TCS was degraded below its detection limit. Additionally, if an aromatic molecule carries an aliphatic side chain, •O- attacks there by H abstraction whilst •OH adds preferentially to the aromatic ring (applied to TCS in its ionized: EBR experiments with 7 and 14 mA/cm$_2$) (Buxton et al., 1988). Performing a statistical comparison with a one-way *t*-test, with 95% confidence, for the highest current densities, in case 1 (Ti/MMO as anode at 14 mA/cm$_2$) and case 2 (Nb/BDD as anode at 10 mA/cm$_2$) regarding the removal rates, Ti/MMO is slightly better at the degradation of TCP (< LD and ~91% , with a *p* value = 0.0061). BDD is clearly more efficiently at degrading MTCS (~47% to 84% with a *p* value = 0.015).

According to the above data, the decision for the best current density in the following sections were based on: (1) Ti/MMO anode, 7 mA/cm$_2$ was chosen because either less intermediate peaks (by-products formed after electro-degradation) were detected in the LC/MS analysis or also, in the case of the highest current tested, the ratio between the energy consumption and the degradation rates were similar and less costly; (2) Nb/BDD anode, 10 mA/cm2 was chosen due to the higher degradation observed for MTCS compound.

### 3.1.3 Electrochemical Batch Reactor (EBR): degradation kinetics

Normalized decay of concentration as a function of time over 4 h, for the two electrodes combination at the best current density conditions, are present in Figure S1 on supplementary data. Considering a pseudo first-order degradation for all contaminants, the corresponding law rate can be written as Equation 2:

$$\mathrm{Ln}(C/C_0) = -kt \tag{2}$$

where $k$ (min$^{-1}$) is the apparent constant rate of the reaction.

According to experimental data, Table 4, the EOCs removal follow a pseudo first-order kinetics ($R^2 \geq 0.9$).

From the kinetics obtained it is possible to estimate that the contaminants removal starts immediately after the application of the electric current (see Figure S1 at supplementary data). The fastest rate was achieved with Ti/MMO for TCS and the slowest for MTCS in both electrodes combination (Table 3). Comparing both anodes, the degradation velocity is lower with Nb/BDD that also presents identical degradation rates for all the analytes under study, showing a more constant degradation velocity behavior. The highest adsorption's strength of •OH radicals on each electrode surface (adsorption enthalpy) the lower oxidizing power (Kapałka et al., 2009). Even though, the BDD electrodes are well known by their high oxidizing power (Alfaro et al., 2006; Kapałka et al., 2009, 2008), the presence of chloride species can slow down the electrodes performance in the EOCs degradation, and therefore the indirect oxidation by active chlorine species may not occur (Anglada et al., 2009; Mascia et al., 2007; Pereira et al., 2015; Scialdone et al., 2009).

Following the kinetics, it is important to mention that EOCs degradation was considered as the elimination of parent organic compound. The parent compound loss indicates transformation at an unknown degree, and not necessarily mineralization, where sometimes the by-products can be more harmful than the parent compounds. The Total Organic Carbon, TOC was analyzed for the different electrode combination (Ti/MMO anode, 7 mA/cm$_2$ and Nb/BDD anode, 10 mA/cm$_2$) in both EBR and EFR treatments. The potential to mineralization is given by comparison data from the TOC in the initial inlet (effluent spiked with EOCs) and the TOC data of the four experiments. Therefore, the TOC decay when Ti/MMO is used as working anode with 7 mA/cm$_2$ in EBR and EFR was 36% and 27%, respectively, in effluent. If Nb/BDD is used as working anode with 10 mA/cm$_2$ in EBR and EFR was 30% and 22%, respectively in effluent. Accordingly, the Ti/MMO as anode seems to present more potential to the full mineralization of the compounds under study. Wachter et al (2019) reported that, for a complete TOC removal many oxidation steps, higher current densities and longtime treatment should be applied. A non-dependence of the TOC removal rate on the electrodes characteristics may be expected for a system where the organic molecule and its oxidation intermediates are oxidized on the anode surface. The results obtained in the present work were in agreement with those reported by the Medeiros de Araújo et al. (2014) on the mineralization of the dye Rhodamine B and Souza et al. (2016) on the mineralization of the DCP.

In this sense, the experimental samples were analyzed by LC/MS. Comparing Ti/MMO and Nb/BDD as working anodes, the latter presents possible by-products from electro-EOCs degradation pathways, while the former produced no detectable by-products (results not shown). The degradation of TCS in chlorine matrices (effluent initial Cl concentration 488.7 ± 593.2 mg/L) led to the formation of two tetra- and one penta-chlorinated hydroxylated diphenyl ether, as well as 2,4-dichlorophenol. Chlorination of the phenolic ring and cleavage of the ether bond were identified as the main triclosan degradation pathways (Canosa et al., 2005). Free chlorine mediated oxidation of triclosan leads to the formation of chloroform and other chlorinated organics (Fiss et al., 2007; Rule et al., 2005). Therefore, TCS is assumed as the main contributor for the identified by-products, since the other three EOCs are TCS natural by-products or metabolite. Thus, the identified ions masses in case 2 (10 mA/cm$_2$) were found and estimated from the profile isotopes of a mass spectrum: ion 128 with one hydroxyl and one chlorine in the phenolic ring, 161 and 196 ions (from TCS ring breaking or from spiking DCP and TCP), 177 ion with two hydroxyls and two chlorines that are not possible to conclude the position on the ring, and 272 ion corresponding to TCS losing the hydroxyls. Due to the obtained data, a more thorough study is needed to assess the exact mechanisms pathways for the four compounds under study, when an electrochemical treatment is applied.

**3.1.4 Electrochemical Flow Reactor (EFR): emerging organic contaminants degradation at the best conditions**

An EFR reactor, mimicking the secondary settling tank in a wastewater treatment plant was developed and tested. The EFR had a flow rate of 9 mL/min, meaning 55 min of retention time to every 500 mL of a batch of effluent spiked with EOCs. The path of the effluent goes through an internal chamber where the electrical current is being applied, followed by a second chamber, the reactor outlet (Figure 1). Using a flow will have significant effect on the oxidation rates, since it can enhance the mass transport of organic species to the electrode surface where they undergo oxidation mainly by the hydroxyl radicals, and increase the turbulence in the system, that is, at that moment, hydrodynamic.

Table 5 presents the effluent pH, conductivity and voltage before and after the EFR treatment, where there were no statistically significant differences between these parameters, contrarily to EBR treatment (see Table 3). All the parameters remained approximately constant.

Figure 3 shows the removal of the target EOCs. Similar to the EBR, the photodegradation and compound volatilization were not taking place in the degradation process, thus the degradation only occurs due to the direct anodic oxidation or indirect oxidation in the liquid bulk.

Accordingly, in approx. 1 h of electrokinetic treatment, the set-up achieved degradation rates for TCP 61% (± 7%), DCP 41% (± 1%), TCS 87% (± 1%) and, MTCS 41% (± 1%) when Ti/MMO was used as the working anode. The lower rates observed for DCP may be explained by the hypothesis that, due to the TCS degradation path, DCP will be formed when the •OH radicals attached the phenolic ring (Yuan et al., 2013). Comparing the degradations rates in case 1 and 2, there are statistical differences at 95% confidence when different anodes are used: for TCP $p$ value = 0.0196, for DCP $p$ value = 0.0079 and for TCS *p* value = 0.0055. However, for MTCS degradation in both cases 1 and 2, no statistically significant differences were found. MTCS degradation data, points out for the compound's dependency in the matrix composition and treatment's retention time. Regarding the removal rates of TCS, DCP and TCP when Nb/BDD was used as anode in the EFR treatment, some clarifications can be point out: (1) processes such as polymerization of phenolic compounds on the electrode surface can occur, thus decreasing the performance of the electrode; (2) formation of organochloride molecules that are resistant to degradation (Korshin et al., 2006); (3) in alkaline conditions (end pH around 7.1) the Nb/BDD reacts with hydroxide (OH–), which can recombine with OH• to form $H_2O$. At alkaline pH and lower current densities, the abundance of OH– retarded the oxidation of compounds by OH• (Hayashi et al., 2016); (4) additionally, though not observed in current work, at alkaline conditions the surface morphology of Nb/BDD may also change, producing inhibitory conditions for organic oxidation, and ultimately degrading the Nb/BDD surface (DeClements, 1997; González-González et al., 2010; Griesbach et al., 2005). Wachter et al. (2019) also observed a decrease in the removal efficiencies using Nb/BDD as anode, when lower applied current densities (5 and 10 mA/cm$^2$) and pH 10 were combined.

In order to understand the dynamics of the degradation process between the two tested reactor designs, an estimation of the four EOCs was performed in the EBR using the pseudo first-order kinetics for 1h treatment. Thus, the data suggested (estimations values obtained from Figure S1, at supplementary data) that EFR leads to an improvement of degradation for TCP (EBR 45% vs EFR 61%), DCP (EBR 28% vs EFR 41%), and MTCS (EBR 12% vs EFR 41%), when Ti/MMO was used as working anode. In the case 2, Nb/BDD as anode, only MTCS (EBR 20% vs EFR 49%) increased its degradation using a dynamic system. For TCS, the degradation occurs, in both anodes from the first 15 minutes on, however the degradation between the flow vs the batch reactors are not statistically different.

In terms of reactor design, EFR using Ti/MMO anode could be a valid choice to a scale-up reactor for EOCs degradation, both by a financial point of view (Ti/MMO cheaper compared to Nb/BDD), an energy consumption perspective (lower current density, same degradation rates; energy consumption according to Portugal energy price for householders (0.2154€) and the kwh for the EBR and EFR treatment: 0.0014€ for EBR using Ti/MMO anode and 0.0029€ using Nb/BDD anode or for EBR 0.0003€ for EFR using Ti/MMO anode and 0.0009€ using Nb/BDD anode, were calculated), but also due to the faster degradation kinetics (when $t = 1, \mathrm{Ln}(C/C_0) = -k\sim$). Furthermore, according to LC/MS analysis, no detectable by-products were observed with this set-up.

## 4. Conclusion

Applied to a secondary effluent, two different electrochemical reactors (batch and flow) were tested for the degradation of triclosan and its derivative by-products: methyl-triclosan, 2,4-dichlorophenol and 2,4,6-trichlorophenol. The compounds elimination promoted by two different anodes, Ti/MMO and Nb/BDD, in the electrochemical batch reactor and the electrochemical flow reactor was evaluated. In both reactors the best electrode combination was accomplished with Ti/MMO, presenting faster kinetics degradation and less dependency on electrical current, achieving similar eliminations with a cheaper electrode. In the batch reactor at 7 mA/cm2, during 4 h, the degradation rates were below the detection limit for triclosan and 2,4,6-trichlorophenol, and 94% and 43% for 2,4-dichlorophenol and methyl triclosan. In the flow reactor, in a 1 h treatment, the degradation efficiencies varied from 41% to 87% for the four contaminants under study. Electrochemical flow reactor implementation in wastewater treatment plants may be considered as viable option from an operational point of view. The combination of a low current density with the flow, and induced matrix disturbance, increased and speed up EOCs degradation.

**Acknowledgements**

This research was funded by Fundação para a Ciência e Tecnologia through projects UID/AMB/04085/2019, UID/FIS/00068/2019 and PTDC/FIS-NAN/0909/2014. This work has received funding from the European Union's Horizon 2020 research and innovation programme under the Marie Skłodowska-Curie grant agreement No. 778045. C Magro acknowledges Fundação para a Ciência e a Tecnologia for her PhD fellowship SFRH/BD/114674/2016. Paz-Garcia acknowledges the financial support from the Excellence Network E3TECH under project CTQ2017-90659-REDT (MCIUN, Spain), and from "Proyectos I+D+i en el marco del Programa Operativo FEDER Andalucia 2014-2020" – UMA18-FEDERJA-279. Cátia de Almeida Santos and André Nunes Jorge to the help given in the LC/MS by-products identification. Authors thank Ricardo Faria for his help with the reactors drawing and Professor Nuno Lapa, Integrated Member of UBiA-NOVA-FCT, who provided all analysis equipment for effluent characterization. This research was anchored by the RESOLUTION LAB, an infrastructure at NOVA School of Science and Technology.